# Automatic Knowledge Extraction with Human Interface

Denley Lam
*BAE Systems FAST Labs*
Arlington, USA
denley.lam@baesystems.com

Steve Schmidt
*BAE Systems FAST Labs*
Arlington, USA
steve.schmidt@baesystems.com

Patrick Hayden
*BAE Systems FAST Labs*
Arlington, USA
patrick.hayden@baesystems.com

*Abstract*—OrbWeaver, an automatic knowledge extraction system paired with a human interface, streamlines the use of unintuitive natural language processing software for modeling systems from their documentation. OrbWeaver enables the indirect transfer of knowledge about legacy systems by leveraging open source tools in document understanding and processing as well as using web based user interface constructs. By design, OrbWeaver is scalable, extensible, and usable; we demonstrate its utility by evaluating its performance in processing a corpus of documents related to advanced persistent threats in the cyber domain. The results indicate better knowledge extraction by revealing hidden relationships, linking co-related entities, and gathering evidence.

*Keywords—knowledge extraction, human interface, natural language processing, system modeling, model based system engineering, cyber, advanced persistent threat*

## 1. Introduction

System engineering and certification is a significant issue in the Department of Defense (DoD) space (AIA, 2018). The complexity and heterogeneity (e.g., legacy, transition, and cloud components) of systems has led to the *model-based systems engineering* (MBSE) approach, and in particular the use of diagrammatical techniques such as SysML. Consider a realistic but notional medium-sized platform containing 50 top-level systems; comprising development, operation, and maintenance functions, and comprised in turn from 200 subsystems. The platform's documentation encompasses at least 500 documents averaging 250 pages each. Table 1, shows realistic metrics for a traditional assessment process applied to this platform.

This manual assessment illustrates several of the challenges faced in performing typical assessments. The volume of system documentation and engineering data to examine for even average-sized systems can be unmanageable using traditional processes. Generating system maps that adequately represent the multiple fidelity levels necessary to simultaneously achieve system understanding and knowledge of deep system-to-system details is challenging given current manual charting techniques.

To address these challenges and simplify tasks such as knowledge extraction, BAE Systems has developed a document discovery tool called OrbWeaver that utilizes an open source software pipeline at its core to automate the process of manually opening every file to derive system and subsystems. OrbWeaver aims to help a system analyst quickly grasp and build a hierarchical model of the system.

TABLE 1. Intensive Manual Assessment Process

| Step | Manual Approach | Metric |
|---|---|---|
| 1 | **Catalog Available Resources:** Manually open each file for inspection, skimming for most valuable content and making note of overall document summary. | 2-4 weeks |
| 2 | **Map System:** Read content and view diagrams in-depth for understanding. Generate system diagrams showing components, hierarchy and connectivity using manual tools such as Visio or PowerPoint. | 3-4 weeks |

OrbWeaver has shown utility in helping reduce analyst time in analyzing corpora of system documents. OrbWeaver automates many facets of the assessment process, enabling enhancements to productivity. We also explore whether OrbWeaver can facilitate user *transfer learning* (Odom & Natarajan, 2018). That is, we ask the question: can a user without domain expertise improve their domain knowledge via interactions guided through OrbWeaver. The current document assessment process is extremely labor-intensive and time consuming, requiring domain experts and manual tools not suited to complex systems-of-systems architectures typical of platforms.

In this paper, we illustrate OrbWeaver's utility through an extended example focusing on analyzing documents detailing *advanced persistent threat* (APT). The cyber domain is illustrative of systems within systems and the complex relationships between threats and actors. We hope to elucidate the advantages and difficulties in using a semi-automated approach to document learning.

In the rest of this work, we first present OrbWeaver's concept of operations (CONOPS) in section two, and review related work in three. We then focus in four on the design and features of the system. Fifth, an example dealing with APTs and the results. Sixth, presents the performance of the system, and seventh outlines areas for future work. Finally, we make concluding remarks in section eight.

## 2. Concept of Operations

Today, most document analysis is performed by manually opening each file for inspection, skimming for most valuable content and making note of overall document summary. To further exacerbate the issue, most organizations do not have a convenient method to depict and characterize entity relationships across the content of multiple documents. An international survey (Xerox, 2016), reported that 46% of business respondents wasted time every day on paper intensive processes.

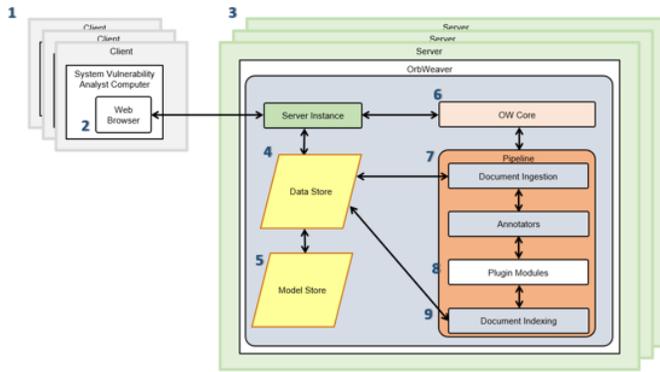

Fig. 1. OrbWeaver Workflow

Government and civilian entities store and maintain large corpuses of documentation and associated media about our complex weaponry systems. These weapon systems are complex assemblies of interconnected devices and subsystems that increasingly rely on computers and software to accomplish their missions. The DOD's system analyst are presented with the challenge of rapidly understanding a weapon system and subsystems without being subject matter experts of that system. They therefore need the ability to rapidly gain this insight and knowledge from the existing system documentation.

OrbWeaver is a scalable and modular document discovery framework for facilitating knowledge based construction (Re et al., 2016) and knowledge graph generation (Ji et al., 2020). The goal of OrbWeaver is to alleviate many of the manual processes associated with document content discovery. In addition, OrbWeaver provides analysts with a graphical environment in which to perform a variety of these tasks at a significant increase in productivity.

Cyber attackers typically target embedded components in these systems to exploit latent bugs and vulnerabilities. OrbWeaver's framework contain additional modules that pertain to unique features associated with security metrics. To understand the risk cyber vulnerabilities present to weapon systems, we must understand the system, subsystems, and connectivity of the system through a document analysis process. Fig. 1 illustrates OrbWeaver's end-to-end workflow, beginning with the analyst interacting with a web browser client through the ingesting, annotation/extraction, indexing, and finally presenting a knowledge graph to the user. Table 2, details the specific steps of the workflow associated with Fig. 1.

### 3. RELATED WORK

The state of autonomy in document management is still in its infancy (Parmenter, 2019). In our survey of related work, we were struck by how much of the data was returned in a raw unstructured format. Tools often exhibited two behaviors; returning too much data that it became burdensome for the user to understand or required specific queries that returned insufficient meaning back to the user.

Document management is recognized to be composed of six broad categories; (i) Automating classification and processing (Nuance Document Imaging), (ii) Data extraction (Google Document Understanding AI), (iii) Clustering documents (iManage RAVN), (iv) Order to unstructured data (GrayMeta AI), (v) Supporting content and document development (Grammarly, Adobe), (vi) Securing documents and data (Adobe). Missing from these six categories is a method of transforming processed data to human understanding. The missing human interface is what OrbWeaver aims to address.

OrbWeaver's focus is on the first four categories of document management which we further refine into *information extraction* (IE) (Niklaus et al., 2018) and *knowledge base construction* (KBC). Underlying IE and KBC is *natural language processing* (NLP), which has quickly evolved from a set of techniques for artisans into a more pragmatic toolkit for dedicated developers (Otter et al., 2019). NLP tools enable the retrieval of syntactic and semantic meaning of a sentence. Towards this end, *parts-of-speech* (PoSs), *named entity references* (NERs), and *word vectors* are commonly extracted features (Sathiyakugan, 2019). Using open-source frameworks such as SciKit, TensorFlow, or Torch, we can apply deep learning to train on big data and generate new inferences. However, while most of these frameworks are accessible to end users, they still remain unintuitive: there is an industry-wide focus on algorithms and less on the human interface.

Google and IBM (IBM Watson Explorer) have been in the forefront for Knowledge Based Construction and Knowledge Graphs (IBM, 2019). In terms of system engineering and system modeling, there exists such tools as Capella Arcadia, Amazon Textract, and MagicDraw (SysML) (No Magic, 2020).

TABLE 2. ORBWEAVER WORKFLOW

| Step | Action Taken |
|---|---|
| 1 | Initial phase of the OrbWeaver process is a client connection to the OrbWeaver server. |
| 2 | The user's browser now has a connection to the OrbWeaver Tomcat server instance. |
| 3 | The OrbWeaver server instance and the open source search platform Solr can be distributed across multiple instances as needed for scalability. |
| 4 | After the client server connection has been built the document pipeline process starts the main component of the OW Core functionality. |
| 5 | The pipeline begins with document ingestion. These data, to include documents and source code are consumed by the data dissection process to generate basic information building blocks called "orbs". |
| 6 | The output from the Document ingestion process is stored in the OrbWeaver data Store. |
| 7 | After ingestion, a sequence of "Annotators," to include Language Modeling, Named Entity Recognition, Acronym Resolution, Co-reference Resolution and Relationship Extraction are executed. These components comprise the NLP portion of the tool. |
| 8 | The output from the Annotators are stored in the model store |
| 9 | The Plugin Module part of the pipeline allows developers to easily add additional features to could include new open source graph algorithms, additional NLP packages or other custom modules. |
| 10 | The final phase of the process involves document indexing. This feature provides a way to easily retrieve any document ingested. |

## 4. DESIGN

OrbWeaver adopts industry best practices and standards by using recognized open source tools for task specific processes. The architecture of OrbWeaver is comprised of two major components; (i) a pipeline feedback loop engineered with open source software for flexibility and scalability. (ii) a custom designed UI to enable user productivity. The underlying combination and presentation of the data in its final form to the user should extend the document management experience beyond automation into human hands.

OrbWeaver supports model based system engineering by allowing the resultant constructed system model to be exported as a SysML file. The SysML file can be imported into other tools to allow for seamless culpability with other existent tool chains. Optionally, exporting as a Microsoft Excel file is supported. In the following sections we describe in more detail the software, algorithms, and user interface underlying our framework.

### 4.1 Open Source Software

The information extraction pipeline consists of seven major open source software products. Apache Nifi is responsible for orchestrating the processing pipeline which consists of a document extractor/parser, a configurable series of annotators, an annotation merger/resolver, and an indexer (Apache Nifi, 2020). An example of an Apache graphical configurable pipeline is shown in Fig. 2. A list of the open source software used is provided in Table 3 (Apache Solr, 2020) (Apache Tomcat, 2020) (Apache Tika, 2020) (Apache OpenNLP, 2020) (Eclipse, 2020).

### 4.2 Algorithms

OrbWeaver uses well-known NLP algorithms, in addition to combination of search and sorting algorithms in its annotation stage of the pipeline. See Table 4 for algorithm pipeline. Word embedding's are representations of text as vectors in a lower dimensional space than traditional one-hot encoding space. This compactness allows for similar words to be mapped into similar regions of the vector space in an effort to capture the semantic context. OrbWeaver leverages a pre-trained Word2Vec model on Google News data (3 million words and phrases as 300-dimensional vectors) (Google, 2013) which outputs, and thereby adds to our data store, word and phrase distance vectors. These enrichments could be furthered modeled for specific domain expertise output or generically clustered for manual consumption.

TABLE 3. OPEN SOURCE SOFTWARE

| Software | Description |
|---|---|
| Apache Tika | Content extraction, detection and analysis |
| Apache Solr | Document-based data store with non-relational data model |
| Apache Nifi | Web-based graphical system to process and distribute data; automated data flow between systems. |
| Apache Tomcat | Java-based web application server |
| Apache NLP | Commercial industry java-based NLP |
| Stanford NLP | Academic java-based NLP |
| Java DL4j | Learning and Prediction: Neural nets, Word2Vec |

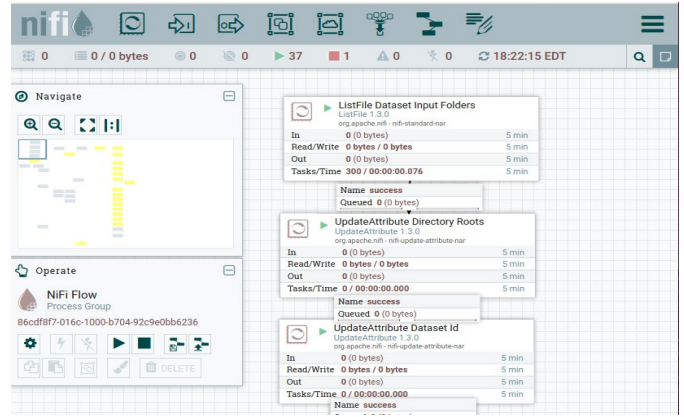

Fig. 2. Example of a Apache Nifi Workflow

OrbWeaver adds an additional layer of enrichment for querying (searching) the created vector space, through the PageRank algorithm (Roberts, 2016). The PageRank algorithm provides link analysis of the underlying graph created by the parsing and node annotation of the Stanford CoreNLP tool (Stanford, 2020) and edge weightings of the aforementioned similarity measures. This allows an interface abstraction to the user to query specific *named entity references* in the graph and obtain a parameterized subgraph of most closely related words. Furthermore, this subgraph contains validation evidence within the GUI in the form of the original sentence text. This allows an operator to explore the graph and conceptually (as of time of writing) interact with the graph weights and edge connections as appropriate for their task or injection of domain expertise.

TABLE 4. ALGORITHM PIPELINE

| Function | Description |
|---|---|
| Document Dissection | Provides the reader / parser functionality for a wide variety of document types. |
| POS Tagging | Annotates words with the appropriate part-of-speech. |
| Grammar Parse | Provides a grammatical parse tree of every sentence. |
| Named Entity Recognition | Automatically identify named entities. A named entity can be a person, device, place, date, number, position, or a number of other categories. |
| Co-reference Resolution | Resolve references (pronouns, titles, indirect objects, etc.) to the subject to which it refers. |
| Relationship Extraction | Extract two types of relationships: co-occurrence and subject-verb-object triplets. |
| Acronym Expansion | Automatically detect acronyms and expand them to the most likely fully qualified phrase. |
| Image Classification | Classify images embedded in documents into a set of broad categories (chart, image, OV1, etc.) |
| OCR | Extract text from embedded images |
| Document Hierarchy | Create a hierarchy of sections within a document using the table of contents. |
| Language Modeling | Converts words to vectors. Decomposing the uses of a word into high dimensional space, reasoning algorithms can more easily operate on them. |

## 4.3 User Interface (UI)

OrbWeaver's NLP annotation pipeline builds semantic linked relationships while the user interface enables human guided domain knowledge interaction. OrbWeaver contribution to knowledge acquisition is its dynamic, functional, and accessible user interface. The objective of this task was to develop a graphical user interface that would provide intuitive control over each OrbWeaver functionality. We sought to make the tool easier to use and more accessible for analysts with a range of expertise levels. Within the GUI we integrated a visualization environment that provides improved user interactivity, scalability, and representation for hierarchically organized systems.

A web based user interface was selected for this purpose. Fig. 3, shows a screen capture of the interface for the OrbWeaver system. The user interface makes extensive use of open source JavaScript libraries such as Angular.js and Bootstrap.

We bridge the document discovery knowledge gap using our OrbWeaver framework to encompass semi-automated learning, modular framework, and big data information extractor within a human friendly interface. Table 5 provides a list of features supported by the UI.

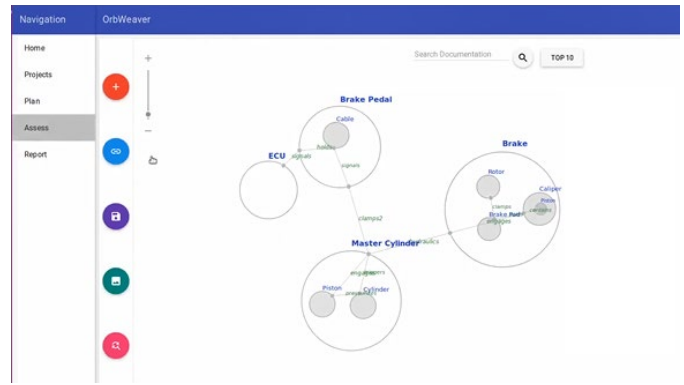

Fig. 3. OrbWeaver Web Interface

TABLE 5. USER INTERFACE FEATURES

| Function | Description |
|---|---|
| Create/Edit/Delete Orb | Creates a default orb. Change the attributes of a given orb. Delete a selected orb. |
| Create/Load/Edit/Save/Delete Project | Define a container for keeping track of the system view for a given dataset. Populate the system view for previous project. Edit and save properties of a project. Remove a project. |
| Zoom/Pan | Zooming in an orb reveals additional details. Pan will quickly place focus and zoom in on a specific orb. |
| Layout Manipulation | Visualization supports automated layouts with the nested force directed graph layout. Manual layout is available with enforced rules. |
| Orb Visual Properties | Orb sizing, colors, border, thickness, and pattern are configurable through a configuration panel. |
| Ingest Documents | Select, upload, and ingest documents with a progress indicator. |
| Show Top Entities | Show a ranked list of the top entities |
| Add/Edit/Delete Links | Add, edit, and delete links between orbs. |
| Show Evidence | Show sentences and documents associated with link. |
| Show Similar Orb | Show a list of entities related to selected orb. |
| Assign Parent and Child Orb | Support the nesting of parent/child orbs. |
| Modify risk profile | Modify risk profile to support susceptibility, credibility, criticality, and resilience. |
| Find and Replace Orb Entities | Find and replace underlying entities and re-generate relationships. |
| Generate System Report | Export system orb view as a SysML or Excel document. |

## 5 EXAMPLE - APTNOTES

We are interested in exploring whether OrbWeaver can provide useable information transfer, as well as deriving knowledge graphs. We selected the APTnotes's GitHub (Bandia, 2020) repository to serve as the corpus for ingestion. APTnotes is a curated list of vendor documents that pertain to cyber-attacks since 2008, malicious activities are associated with Advanced Persistent Threat (APT) groups.

OrbWeaver ingested 506 documents that contain XLSX, CSV, and PDF files. In order to provide ground truth, we utilized the Cyber Analytic Repository (CAR) a tool created by MITRE and visualized with their CARET (The CAR Exploration Tool) (Mitre, 2020 a) system. CARET visualizes APT groups and their association with the ATT&CK knowledge base. MITRE's ATT&CK maps real world attack tactics and techniques for the development of threat models and methodologies (Mitre, 2020 b).

Fig. 4, show different APT groups using different techniques to gain initial access, perform collection, and execution. MITRE's ATT&CK filters APT by groups; we explore the three groups Carbanak, Gothic Panda, and Dark Caracal. (i) Carbanak a threat group that traditionally targets banks using remote desktop software. (ii) Gothic Panda a China-based threat group targeting US and political organizations in Hong Kong. (iii) Dark Caracal a Lebanese threat group specializing in phishing documents.

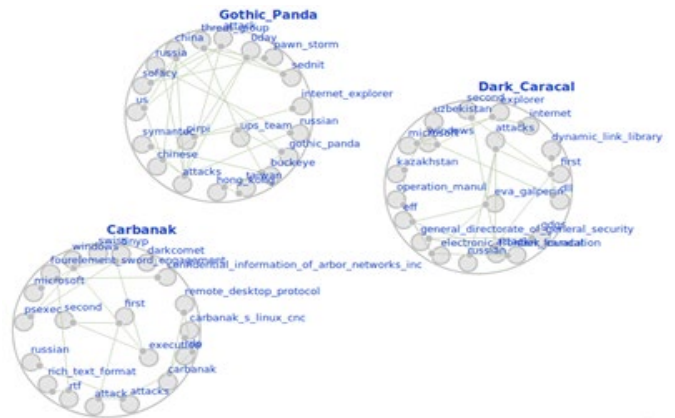

Fig. 4. APT group orbs generated by OrbWeaver

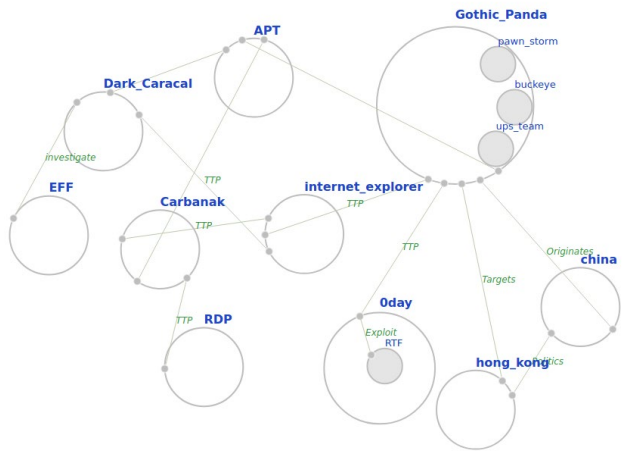

Fig. 5. APT user drawn model with OrbWeaver

OrbWeaver assumes the analyst has rudimentary knowledge of the domain. A higher expected utility is observed when the analyst has basic familiarity with the documents ingested. Without the basic knowledge from reading the ATT&CK group summary, the nodes require more work to decipher.

OrbWeaver's knowledge graph strengths are displayed twofold. Firstly, OrbWeaver highlights important named entities and extracts additional relationships between correlated documents that are not inherently obvious. Comparing OrbWeaver's graphical orb output to the ATT&CK group, we observe comparable nodes in additional to uniquely discovered node linkages.

The Dark Caracal node links to the general directorate of general security (GDGS) node, the GDGS is a Lebanese threat group. This is a hidden relationship reference until we reveal the evidence of the linkage. However, the ATT&CK group summary did not mention the uniquely linked "operation manul" and Kazakhstan nodes.

Secondly, OrbWeaver reveals its strength in pulling together disparate documents with comparable semantic meaning. OrbWeaver presents the most important entities in the orbs without overwhelming the analyst with the minutia of all the tactics and techniques (TTP), see Fig. 5.

The analyst can quickly scan the three APT groups for similarities and differences within its visual interface. Observing that Carbanak uses mainly remote desktop protocol, RTF phishing, and psexec. These TTPs differs from Gothic Panda which uses 0day, internet explorer, and targets Hong Kong. Again, distinctly, Dark Caracal uses DLL, internet explorer, and operates mostly near Russia. The similarities are minimal, besides Windows and Russia.

Using OrbWeaver's top ten results, we observe that Russia, China, and Microsoft are the most cited terms. The analyst at this point may choose to organize the orbs by geographic region, TTP, or threat actor. Interacting and drawing with the orbs will allow the analyst to generate a customizable system model of the documentation.

## 6 PERFORMANCE

In order to facilitate ease of testing, we installed OrbWeaver on our test laptop inside of a Virtual Machine (VM) that supports two CPUs at 2.8 GHz and 32GB of RAM. We observe from the benchmarks of our open source software that the bottleneck lies in the NLP processing, see Table 6. We predict that by scaling horizontally and vertically, processing times will approach the higher performance number supported by the Nifi pipeline. Ideally, OrbWeaver would run once on a corpus of documents and would not overlap into analysis time. We extrapolate that worst case scenario, 1 TB of data will take 1 week to process using 4 rack-mounted servers (Stelly, 2017). Additional testing with GPU's may yield significant performance improvement, but was not tested.

TABLE 6. EXPECTED SOFTWARE PERFORMANCE

| Software | Metric |
| --- | --- |
| Apache Tika | 1 TB of data using 192 cores is 24 hours |
| Apache Nifi | 100 MB/s, limited by JVM |
| Apache NLP | 20 sentences/200 token with an average 3.7 seconds |
| OrbWeaver | .03 MB/sec with test VM |

## 7 FUTURE WORK

Future development plans that would be proposed with proper funding include extending OrbWeaver through the use of third party natural language processing tools to perform automatic discovery of system components, determine their descriptions, and identify relationships they share with other components. These findings will be suggested to analysts who can then choose to instantiate them in the system map. When a system element is created, it is automatically linked with the relevant information elements and documentation resources that describe it for quick future reference and easy pivoting to related systems. Fig. 6 illustrates three different depth levels that can be configured during the construction of the knowledge graph with varying complexity.

Leveraging leading third party named entity and relationship extraction libraries will increase OrbWeaver's autonomy in recognizing systems and information pertaining to systems. We will tailor these algorithms or re-train them as necessary to locate systems related elements. We will incorporate capabilities that enable recognition of binary files and sources that correspond to different system nodes, and automatically associate them with the system model.

System level cyber assessments can often take many months, even years in some cases, to complete. In one organization's Cyber Risk Assessment Framework, many of the manual tasks of the System Architect naturally align to automation using OrbWeaver.

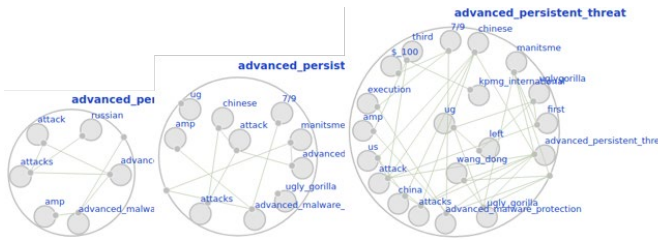

Fig. 6. Three different configurable depth and complexity for the same node in a knowledge graph

A key aspect of the cyber assessment requires the team doing the assessment to determine the system attributes and data requirements. The data requirements listed include architectural documentation, system interface documentation, H/W and Software (S/W) information, H/W and S/W configurations, technical or maintenance documentation, mission information, traditional FMECA and Mission Essential Subsystem Matrix (MESM) information. All of these supporting information could be ingested into OrbWeaver and the corresponding system and subsystem diagrams could be generated to assist in the Cyber Risk Assessment.

## 8 Conclusion

Manual document analysis is often time consuming and prone to error. As the volume of data that organizations are required to keep on hand continues to grow, having tools that will help automate the discovery of valuable content from these data is critical.

Most government agencies are now mandating a system's level of cyber vulnerability to assess the current level of cyberspace threat to the system. Many agencies are providing thorough guidance and methodologies to conduct these assessments. However, there is a lack of adequate tools to automate many of the manual steps in the process.

OrbWeaver can aid in content discovery and automated system level analysis across multiple documents sources. Its graphical environment enable the analyst to interact with the extracted content and to perform a variety of these tasks at a significant increase in productivity. OrbWeaver has a modular and scalable system architecture paving the way for future upgrades and demands.

### Acknowledgment

This material is based upon work supported by The Naval Air Warfare Center under contract N68335-16-C-0157. Any opinions, findings, and conclusions or recommendations expressed in this material are those of the author(s) and do not necessarily reflect the view of the Naval Air Warfare Center, Department of Defense or the U.S. Government, (Public Release # 2020-628). In addition, funding for supplemental research and development was provided by BAE Systems.

The author(s) gratefully acknowledge all contributors past and present to the project.